\documentclass[newstyle,twocolumn,proceedings]{rmaa}

\usepackage{rmaacite}

\renewcommand{\P}[1]{%
\ifnum#1=1\hbox{OW~168--326E}\fi
\ifnum#1=2\hbox{OW~167--317}\fi
\ifnum#1=3\hbox{OW~163--317}\fi
\ifnum#1=5\hbox{OW~158--323}\fi
\ifnum#1=0\hbox{OW~171--334}\fi}

\title{The ionization of the emission line gas in nearby powerful radio galaxies}
\author{Clive N. Tadhunter 
  \affil{University of Sheffield} }

\fulladdresses{
\item C. N. Tadhunter: Department of Physics \& Astronomy, University of Sheffield, Sheffield S3 7RH (c.tadhunter@sheffield.ac.uk).}

\shortauthor{Tadhunter}
\shorttitle{Ionization mechanisms in radio galaxies}


\abstract{
In this article I review the dominant physical mechanisms in the 
extended emission line regions (EELR) of powerful radio galaxies 
at low redshifts and, in particular,
the balance between jet-induced shocks and AGN photoionization.
I consider the evidence for jet-induced shocks based on 
morphological, kinematical and ionization (diagnostic diagram) information.
Although each of these three types of information separately provides some evidence for the effects of jet-induced shocks, this evidence is often ambiguous. The major advance in recent years has been to combine 
morphological, kinematical and diagnostic diagram approaches. In this way it has been
possible to show that jet-induced shocks dominate the ionization and acceleration
of the EELR along the radio axes of many powerful radio galaxies. However,
AGN photoionization is likely to remain the dominant mechanism in the nuclear
regions of radio galaxies in which the radio sources extend well beyond the
emission line regions.}   
\listofauthors{C.~N.~Tadhunter}
\indexauthor{Tadhunter, C.~N.}

\begin{document}

\maketitle

\section{Introduction}
\label{sec:intro}
The
extended emission line regions (EELR: $1 < r < 100$kpc) around powerful
radio galaxies have the
potential to provide key information about the nature of the
central energy sources (e.g. \pcite{rawlings91}), the
triggering and fuelling of the activity (e.g. \pcite{tadhunter89c},
\pcite{baum92}), and the evolution of the host galaxies. However,
if we are to use them in this way, it is crucial to understand the
dominant physical mechanisms. In particular: to what extent are the observed
emission line features a consequence of the activity, and to
what extent do they reflect the intrisic (pre-activity) properties
of the host early-type galaxies? 

In this review I will concentrate on the issue of the ionization mechanism
for the emission line gas, which has been a perrenial problem for
all types of active galaxies since they first started
to be studied in depth more than 50 years ago. Like so many fields of
astrophysics, this is an area which has benefitted from the considerable
improvements in instrumentation and detector technology in the last two decades. At
optical wavelengths is now possible to combine high spectral resolution and
sensitivity, with wide spectral and spatial coverage. Reflecting this, most
of the recent advances in this field have come about by combining ionization,
kinematic and spatial information.

I will start by reviewing the main models that have been proposed
to explain the extended emission line properties of powerful radio galaxies
at low redshifts ($z < 0.4$).
I will then consider separately the approaches of using morphological,
kinematical and emission line ratio information to determine the dominant ionization mechanism,
before demonstrating the power of combining these approaches. 

\section{Models: AGN illumination vs. jet-induced shocks} 
\label{sec:models}

Many physical mechanisms can potentially influence the properties of the
extended ionized gas, but the two which have received most attention in recent
years are: anisotropic AGN illumination and jet-cloud interactions. 

The {\it anisotropic illumination} model is suggested by
anisotropy-based unfied schemes, which propose that radio galaxies and
quasars are the same thing viewed from different directions, with the
quasar nucleus blocked from our direct view in the radio galaxies by a central
obscuring torus (e.g. \pcite{barthel89}). If the unified schemes
are  correct, most radio-loud
active galaxies should have a powerful source of ionizing photons (the
quasar), which illuminates the ISM of the host galaxies in a bi-cone pattern. 
It is clear that this mechanism {\it must} be important at some level,
because optical polarization studies reveal reflection
nebulae around several powerful radio galaxies, and  spectra of the
polarized light show broad quasar-like emission lines (e.g. \pcite{cohen99})
--- properties which cannot be explained in any other way than by anisotropic AGN illumination. 
This mechanism also appears energetically feasible in the
sense that covering factors of, typically, only a few percent are required to
produce the observed total emission line luminosities (10$^{42} < L_{em}^{tot} 
< 10^{45}$ erg s$^{-1}$) by quasar
photoionization, given that quasars have ionizing luminosities in the 
range 10$^{44} < L_{ion} <
10^{46}$ erg s$^{-1}$. The major predictions of this model are: broad emission line
distributions which reflect the wide opening angles of the quasar illumination cones ($\delta \theta \sim$90 -- 120$^{\circ}$); quiescent,
gravitationally-induced
emission line kinematics; and a wide range of ionization states, reflecting
the range of possible ionization parameters, cloud optical depths, and
ionizing contiuum shapes.

The main alternative, often labelled the {\it jet-cloud interaction} model, actually
encompasses a range of situations in which the radio-emitting plasma 
interacts with the ISM.
These include: direct jet-cloud interactions, entrainment of material in the
turbulent boundary layers of the jets, and interaction between the ISM
and the bow shocks driven by the expanding
radio lobes and hot spots. In general, the interaction between the
radio-emitting plasma and the ISM is likely to be complex, but at the very
least we would expect the warm emission line gas to be accelerated,
compressed, heated, and ionized by shocks driven by the relativistic plasma.
Energetically, there is little to choose between this model and AGN
illumination, since, for typical bulk jet powers, the conversion efficiency
required in individual sources to explain the emission line luminosity entirely
in terms of jet energisation is $\sim$0.5 -- 5\% (\pcite{clark96})
--- similar to the covering
factor required in the AGN illumination model. However, the
jet-cloud interaction model predicts quite
different properties for the emission line gas. These include: close, detailed
morphological associations between radio and optical emission line features; 
relatively high electron densities and temperatures, due to the compression
and  heating effects of the shocks; and a wide range of ionization, depending
on the shock speed, magnetic parameter, and balance between precursor and
cooling post-shock gas (e.g. \pcite{dopita96}).

\begin{figure}
  \includegraphics[width=8cm,angle=0]{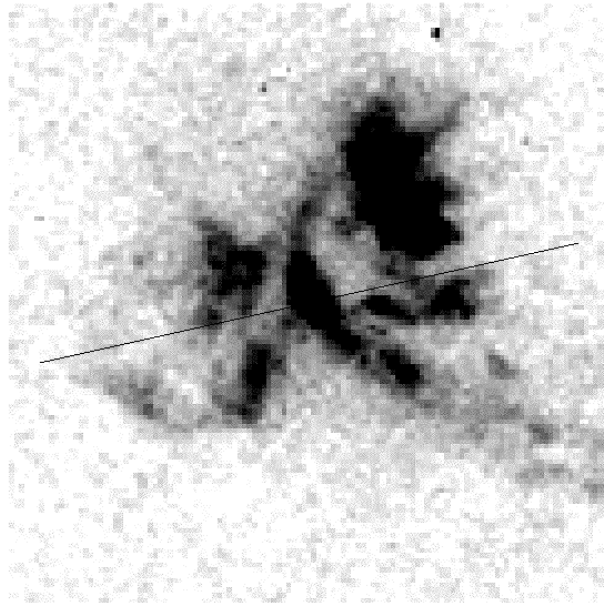}
  \caption{Narrow-band H$\alpha$ image of the central regions of
the powerful radio galaxy Cygnus A ($z = 0.0567$) taken with WFPC2 on the
HST (see \pcite{jackson98} for details). North is to the top, east to left,
and the image is centred on the nucleus. The line segment shows the 
approximate direction of the radio jets. Note the
cone-like morphology on the NW side of the nucleus.}
  \label{fig:cones}
\end{figure}

It should be emphasised that the choice between these mechanisms is unlikely
to be
``either/or'';  
it is probable that both mechanisms make a significant contribution. Because
of the $1/r^2$ geometric dilution of the ionizing radiation field, the AGN illumination 
is most likely to dominate close to the nucleus ($r< 10$kpc, although this
depends on the radial variation in the covering factor). On the other hand,
the radio jets can project their power on radial scales of 100's of kpc. Thus,
jet-cloud interactions are more likely to dominate on a larger scale.

It is also important to bear in mind that AGN illumination and 
jet-cloud interactions are not the only mechanisms that
are capable of affecting the emission line properties. For example,
given the presence of energetic cosmic rays associated with the radio-emitting
plasma, cosmic rays may provide a significant ionizing input \cite{ferland84}. 
Moreover, quasars and/or circum-nuclear starbursts can drive
winds which are capable of accelerating the emission line clouds to large
velocities (e.g. \pcite{heckman90}) --- a possible alternative to acceleration by jet-induced
shocks. However, whilst noting that these alternative mechanisms may be
important, I will not consider them further in this review.

\section{The Evidence}
\label{sec:evidence}

I will now discuss the extent to which the predictions of the two main 
models are consistent with the morphological, kinematical and ionization
properties of the EELR in powerful radio galaxies.

\subsection{Morphologies}
\label{sec:morph}
One of the most striking results to emerge from studies of the optical/UV
morphologies of radio galaxies is the close alignment
between the radio and optical emission line axes: the so-called 
``alignment effect''. Although this alignment effect appears strongest  
in  high redshift objects \cite{mccarthy87}, a significant, though
looser, correlation between radio and optical emission line axes
has also been found in lower redshift objects \scite{baum89}.
Unfortunately, the alignment effect does not, by itself, allow one to
distinguish the dominant physical mechanism, because significant alignments
are predicted in both models. In order to distinguish the dominant
physical mechanisms it is necessary to look at the morphologies in
greater detail.

First, I consider low redshift radio galaxies ($z < 0.4$). In most cases the
distribution of emission line in the low-z objects is broad in the sense
that the EELR cover a wide range in azimuthal angle \cite{baum88}.  Allowing for the
fact that the axes of the ionization cones may be tilted relative to the plane
of the sky, such broad distributions are consistent in most cases with AGN
illumination (e.g. \pcite{fosbury89}).
However, sharp-edged ionization
cones of the type visible in some Seyfert galaxies (e.g. \pcite{pogge88}, \pcite{tadhunter89c}), appear to be rare in powerful radio galaxies; 
currently, the best example is found in the near-nuclear
regions of the nearby powerful radio galaxy Cygnus A (\pcite{jackson98}, see Figure 1). The
apparent absence of well-defined ionization cones is likely 
to be due to the fact
that the warm ISM in the host
galaxies is clumpy, irregular  and  has a low filling factor.
Thus, while the emission line morphologies of low-z radio galaxies are consistent with the aniostropic illumination model in many cases, they
cannot be said to provide strong evidence to support this model.

The morphological evidence for jet-cloud interactions  is 
more compelling. A small but significant subset of low redshift 3C radio 
galaxies ($\sim$10 -- 20\%) shows detailed morphological associations between
radio and optical features. A particularly good example, which illustrates
the range of phenomena associated with the jet-cloud interaction mechanism,
is provided by the 100 kpc-scale emission line nebula around Coma A
(\pcite{vanbreugel85}, \pcite{tadhunter00}). In low
signal-to-noise emission line images, the nebula in this object appears as a series of 
high surface brightness knots which are closely aligned along the radio axis. Indeed, the radio jet appears to be deflected at the site of the brightest of these knots (Figure 2b). However, in deeper tunable filter
emission line images (Figure 2a) a spectacular system of emission line arcs
and filaments is revealed, which circumscribes
the radio lobe on the north side of the nucleus. Most plausibly, the radio jets and lobes in
this source are expanding into a more extensive halo of warm/cool filaments
associated with an interacting group of galaxies. In this case,
images demonstrate the ionizing effect of the jets and lobes 
on the warm
gas.  Note that the emission line structures in Coma A cannot
be reconciled with the AGN illumination model because some of the arc features
wrap a full 180 degrees around the nucleus.
\begin{figure*}
  \includegraphics[width=9cm,angle=-90]{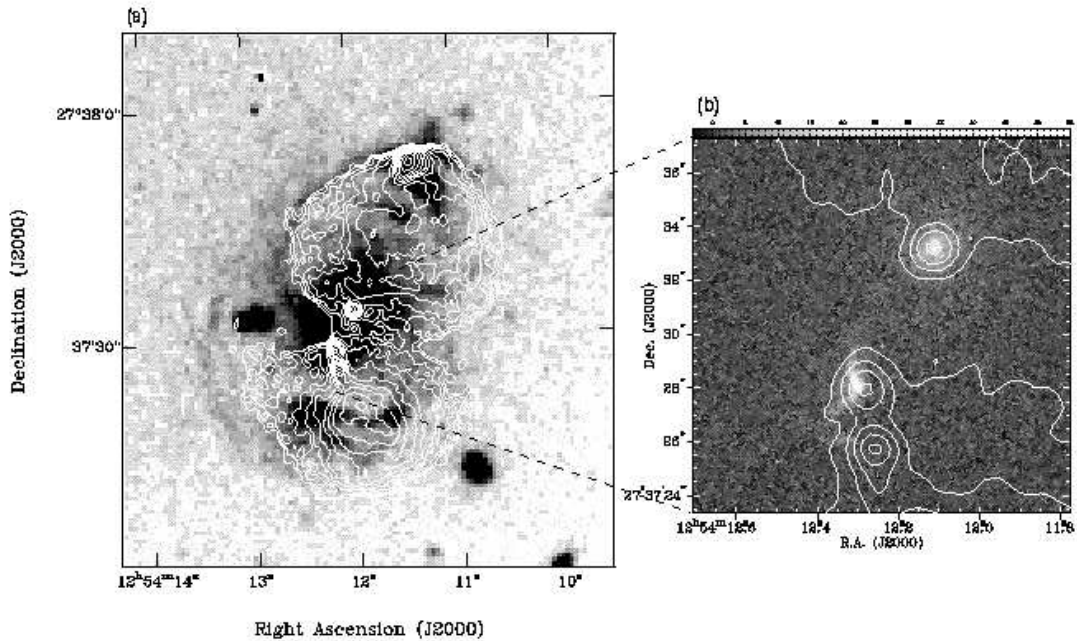}
  \caption{Radio and optical images of the low redshift radio galaxy
Coma A ($z = 0.085$): (a) ground-based H$\alpha$ image (greyscale) taken with the Taurus Tunable Filter on the WHT with the 6cm radio contours superimposed (see
\pcite{tadhunter00} for details); (b)
high resolution [OIII] image (greyscale) of the nucleus and emission line knot to the
south of the nucleus with 6cm radio contours superimposed.}
  \label{fig:comaboth}
\end{figure*}

There also exists strong morphological evidence for jet-cloud interactions
in high redshift ($z > 0.6$) 3C radio galaxies. Not only are
optical/UV structures in such sources closely aligned with the 
radio axes, but in $\sim$30 -- 40\% of cases HST images
reveal that the optical/UV structures are also {\it highly collimated}, with
a jet-like appearance (e.g. \pcite{best96}, \pcite{mccarthy97}). Such highly collimated structures
are difficult to reconcile with the broad ionization cones predicted by
the unified schemes. Furthermore, the most highly collimated optical
structures are associated with sources in which the radio and optical 
structures have a similar spatial extent. Taken together, 
this evidence suggests that
the jet-induced shocks may dominate the ionization of aligned structures
in many of the high redshift sources.

One outstanding issue concerns the extent to which the highly collimated
emission line
structures observed in some high-z sources reflect the intrinsic distribution of
warm/cool ISM in the host galaxies, and the extent that they reflect 
the ionization pattern induced by the radio jets and AGN. While it
is difficult to rule out the idea that the gas is intrinsically
aligned along the radio axis (e.g. \pcite{west94}),
deep images of Coma A (Figure 2a) 
and other low-z radio galaxies indicate that the distribution of warm ISM
is more extensive than might be suggested by
the high surface brightness structures aligned along the radio axis.
Therefore, I expect future deep emission line
images to reveal extensive haloes of the warm ISM
well away from the radio axes of the high-z sources.     

\begin{figure*}
  \includegraphics[width=9cm,angle=-90]{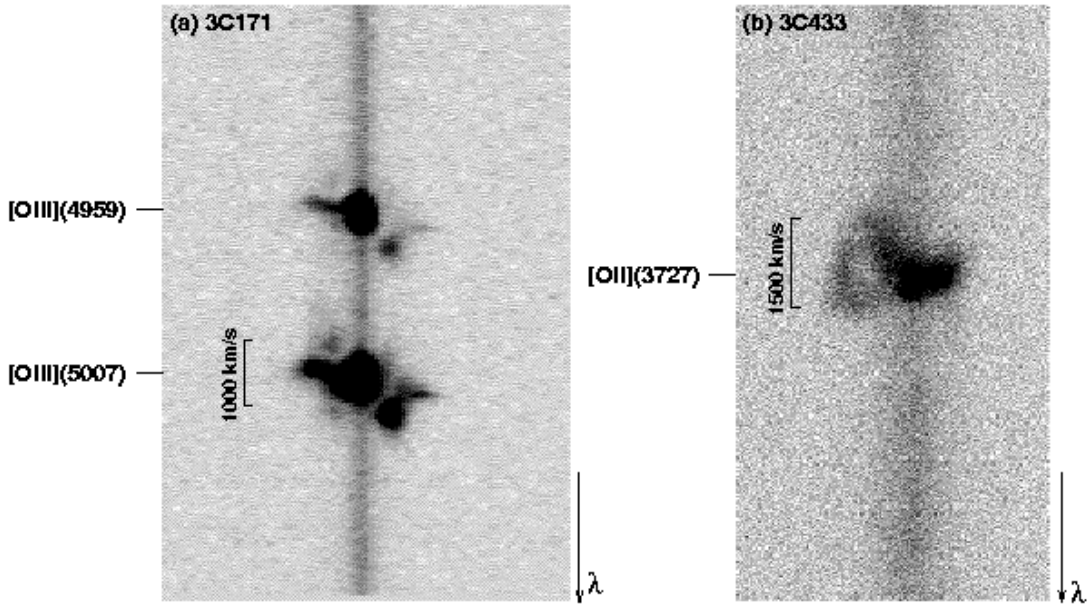}
  \caption{Long-slit spectra showing extreme emission line kinematics
in (a) 3C171 ($z = 0.238$, \pcite{clark98}), and (b) 3C433 
($z = 0.102$). These data were taken
with ISIS on the WHT. For 3C171 the spectrum was taken with the slit
aligned along the radio axis, whereas for 3C433  the slit
misaligned by $\sim$50$^{\circ}$ with respect to the radio axis.}
  \label{fig:kin}
\end{figure*}

\subsection{Kinematics}
\label{sec:kin}

The kinematical properties of the emission line nebulae appear
to correlate broadly with their morphological properties. 

In most low redshift radio galaxies in which the EELR are not closely
associated with the radio structures, the emission line kinematics are quiescent,
with relatively low velocity amplitudes ($\Delta V < 700$ km s$^{-1}$)
and small line widths (FWHM $ < 600$ km s$^{-1}$) that are consistent with
gravitational motions in the host galaxies (\pcite{tadhunter89c}, \pcite{baum00}). 
However, in at least some of the low redshift sources with
closely aligned optical and radio structures, the kinematics 
are more disturbed. One of the most spectacular examples of such
disturbed kinematics is found in the intermediate redshift radio
galaxy 3C171 \cite{clark98}. This object shows line splitting ($\Delta V \sim
1200$ km s$^{-1}$), an underlying broad emission line component 
(FWHM $ \sim 1400$ km s$^{-1}$), and a central narrow component
(FWHM $ < 500$ km s$^{-1}$)
that are detected across the full extent of the high-surface-brightness emission line structure along the radio axis, and on both sides of the nucleus
(Figure 3a). Such
extreme kinematics are
difficult to explain in terms of gravitational motions in the host potential,
but are entirely consistent with the effects of shock acceleration,
especially when the effects of
the entrainment of warm clouds in the host post-shock gas are taken into account
(\pcite{villarmartin99}).

There is evidence for a significant increase in both
the linewidths and the velocity half-amplitudes in 3C radio galaxies
beyond a redshift of $z=0.6$ \cite{baum00}. One explanation for
this trend 
is that the environments of the more powerful, higher
redshift sources are richer than those of their lower redshift counterparts, so that their velocity amplitudes
reflect graviational motions in  the cluster potentials rather than in the 
host galaxy potentials \cite{baum00} While this explanation may
hold in some cases, it is unlikely that it can account for the disturbed
emission line kinematics and large line widths measured in many
of the high-z sources with highly-collimated optical/UV structures
\cite{best00}. 
Such sources
are similar to 3C171 (Figure 3a), and jet-induced shocks are a more 
plausible acceleration mechanism for the warm gas (Best, these proceedings).

Despite the general tendancy for the most extreme emission line
kinematics to be found in EELR that are morphologically associated with radio structures, there
are some important exceptions.
For example, 3C433 (see Figure 3b) exhibits some of the most 
extreme off-nuclear emission line kinematics measured
in a low redshift radio galaxy, with velocity splittings of up to
$\sim$1500 km s$^{-1}$, yet its emission line structures are not aligned
along the radio axis and fall outside the extended radio lobes. On
the other hand,
the bright emission line knot to the south of the nucleus in Coma A is clearly
undergoing a strong jet-cloud interaction (see Figure 2b), but its emission
line kinematics are relatively quiescent, with no clear sign of line broadening
in the existing spectra (\pcite{vanbreugel85}, \pcite{clark96}).
These results serve to emphasise that
it can be difficult to distinguish the dominant physical mechanism(s) purely
on the basis of emission line kinematics.

\subsection{Diagnostic diagrams}
\label{sec:ion}

Since the pioneering work of \scite{baldwin81}, a popular
way of distinguishing the dominant ionization mechanism has been to use 
diagnostic diagrams in which one emission line ratio is plotted
against another, and the results are compared with the model predictions. With
current observational techniques it is now possible to measure several
faint diagnostic emission lines to high accuracy, and to plot a range of 
diagnostic diagrams. The ionization models have also grown increasingly
sophisticated. Not only have shock models been calculated which include both  
post-shock (cooling zone) and precursor emission components for a range
of shock velocity and magnetic parameter (Dopita \& Sutherland 1995, 1996), but
AGN photoionization models have also
been produced which combine a mixture of matter-bounded (optically thin)
and radiation-bounded (optically thick) components \cite{binette96}.

Figure 4 shows a selection of diagnostic diagrams in which various ionization models are compared with the optical
emission line ratios for extra-nuclear emission line regions in
radio galaxies. The models plotted are: standard, optically
thick, single slab, power-law photoionization models (e.g. \pcite{robinson87});
mixed-medium photoionization
models \cite{binette96}; pure post-shock models, and post-shock$+$shock-photoionized
precursor models \cite{dopita96}. All the models assume solar abundances and
low densities. 
Unresolved nuclear emission line regions are not
included because their ratios may be affected by density
stratification effects close to the central AGN.
\begin{figure*}
  \includegraphics[width=17.5cm,angle=0]{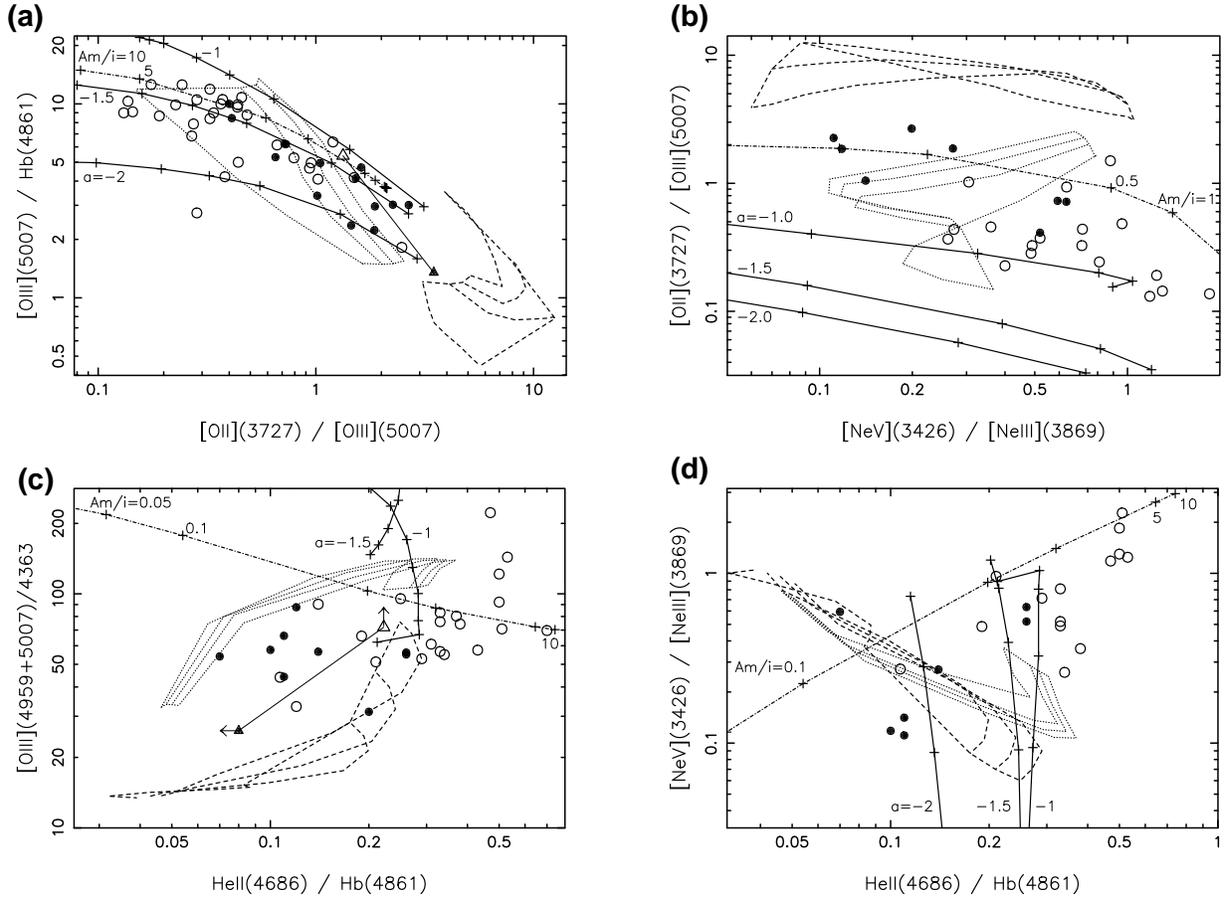}
  \caption{Diagnostic diagrams for the following line ratio pairs: a)~[OIII](5007)/H$\beta$~vs~[OII](3727)/[OIII](5007); b) [OII](3727)/[OIII](5007) vs [NeV](3426)/[NeIII](3869); c) [OIII](4959+5007)/4363 vs HeII(4686)/H$\beta$, and d) [NeV](3426)/[NeIII](3869) vs HeII(4686)/H$\beta$. The emission line data for jet-cloud EELR (filled symbols) and non-jet-cloud EELR (open symbols) have been taken from the following references: \scite{tadhunter86}, ss\cite{tadhunter94}, \cite{storchi96}, \scite{carmen2001}, \scite{clark96}, \scite{clark98}, \scite{clark97}, \scite{robinson00}, \scite{villarmartin98}, \scite{villarmartin99}. The triangles
joined by a solid line represent broad (filled triangles) and narrow (open triangles)
kinematic components in PKS2250-41 \cite{villarmartin99}. Pluses linked by solid lines represent line ratios predicted by optically-thick single-slab power-law ($F_{\nu} \propto \nu^{\alpha}$) photoionization models (using the MAPPINGS
code) with spectral indices of $\alpha$ = -1.0, -1.5 and -2.0, and a sequence in the ionization parameter covering the range $5 \times 10^{-4} < U < 10^{-1}$. Pluses linked by a dot-dash-dot line indicate the predictions of the mixed medium photoionization models from \cite{binette96}, with the ratio the solid angles covered by the matter- and ionization-bounded components (A$_{M/I}$) in the range 10$^{-4} \leq$ A$_{M/I} \leq$ 10. U and A$_{M/I}$ increase from right to left in a), and from left to right in b), c) and d). Predictions of pure post shock models (dashed lines), and 50 per cent shocks + 50 per cent precursor models (dotted lines) from \cite{dopita95} and \cite{dopita96} are also plotted, each sequence corresponding to a fixed magnetic parameter (B/$\sqrt{n}$ = 0, 1, 2, 4 $\mu$Gcm$^{-3/2}$) and a shock velocity varying across the range 150 $\leq v_{s}\leq$ 500 km s$^{-1}$).}
  \label{fig:diag}
\end{figure*}

Several points emerge from examination of these diagrams:
\begin{itemize}

\item {\bf Photoionization models}. The mixed medium photoionization
models provide a significantly better fit to the locus of measured points
than the standard, optically thick models on a range
of diagnostic diagrams (see also \pcite{binette96}). In particular, the 
standard model ionization parameter (U) sequence falls almost orthogonal to the locus of points in the [NeV]/[NeIII] vs.
HeII/H$\beta$ diagram, whereas the sequence obtained by varying
the relative
contributions of matter- and radiation-bounded components (A$_{M/I}$) in
the mixed medium model
provides a better, if not perfect, fit
to the main trend.
\item {\bf Shock models}. The shock models which combine precursor and
post-shock components provide a much better fit to the locus of points than 
the models which include only the post-shock component. There is considerable
overlap between the shock$+$precursor models and the photoionization models
in many of the diagrams. This
is as expected given that, at the faster shock speeds, the shock+precursor 
spectrum is dominated by the precursor emission {\it photoionized} by the
hot post-shock gas. 
\item {\bf [OIII](5007+4959)/4363~vs.~HeII/H$\beta$} (Fig 4c). This diagram
is particularly problematic for the standard photoionization models
(\pcite{tadhunter89b}), although the mixed medium models can at
least explain the low values of [OIII](5007+4959)/4363 (and correspondingly
high electron temperatures) measured in some
cases \cite{wilson97}. 
While the shock  models
predict high electron temperatures and small [OIII](55007+4959)/4363 they do no produce a good fit to the overall
locus of points on the diagram.
\end{itemize}
To summarise, none of the models provides a perfect fit to the data
on all of the diagrams. However, in terms of the consistency of the
positions of the points relative to the models on a range of 
diagnostic diagrams, and also in terms of explaining the shape of the
correlations and the spread of the observed points, the mixed-medium 
photoionization models provide the best fit overall. An important lesson
from the success of the mixed-medium models is that, in order to interpret
the emission line spectra, it is crucial to understand the physical state of
the ISM. Images of the systems of emission
filaments
in the halo of  Centaurus A   -- the closest and best resolved radio galaxy --
demonstrate
the considerable complexity of the warm ISM in these galaxies \cite{morganti91}. Clearly,
we can no longer regard EELR as solely comprising an ensemble
of spherical, optically thick
clouds.  

Note that the fact that the fits are not perfect does not necessarily imply that
the basic principles behind the models are wrong. Rather, it is more likely
that some of the assumptions are wrong, or that the models are over-simplified.
There exists considerable latitude to fine-tune
the models to provide a better fit on all the diagrams. For example, in the
mixed-medium model the column depths of the components, the shape of the ionizing
continuum, ionization parameter, and the abundances could all be adjusted in an attempt
to improve the fit. 

Where does all this leave the question of the dominant physical mechanism?
One way to approach to this question to determine whether the EELR with independent evidence for jet-cloud interactions
fall in  
different positions on the diagnostic diagrams to the EELR that show
no such evidence. Although there is some overlap between the two
groups, one clear trend to emerge is that the
jet-cloud EELR (filled symbols) fall towards the lower ionization
end of the sequence, whereas those without evidence for
jet-cloud interactions (open symbols) 
fall towards the higher ionization end. 
Furthermore, although the shock+precursor models do not provide a good fit to
the locus of radio galaxy points as a whole, they provide as good a fit
as the mixed-medium models if we  consider
the jet-cloud EELR alone. Of particular
note is the fact that the simultaneous
measurement of low HeII/H$\beta$ {\it and} low [OIII](5007+4959)/4363  in some jet-cloud EELR
is difficult to reconcile with the mixed-medium photoionization
models, but is predicted by the shock and shock$+$precursor
models. 
Thus,
for the jet-cloud interaction candidates, the line ratio 
data are at least consistent with
the idea that jet-induced shocks have a significant ionizing input, although
again they do not provide conclusive evidence for shock ionization.

\subsection{Combined approaches}
\label{sec:comb}

The morphologies, kinematics and 
diagnostic diagrams each provide substantial evidence to support
the idea that the radio jets and lobes ionize and accelerate the 
EELR as they expand through the haloes of the host galaxies. However, 
none of this evidence is by itself conclusive.

Recently more compelling evidence for the effects of jet-induced
shocks has been obtained by making detailed
studies of low/intermediate  redshift  radio galaxies
which combine the morphological, ionization
and kinematical information (\pcite{clark96}, \pcite{clark98},  \pcite{clark97}, \pcite{villarmartin98},
\pcite{villarmartin99}). The main results can be summarised as follows.
\begin{itemize}
\item {\bf Ionization minima.} Considering the 
spatial variation in line ratios measured
from low resolution spectra, there is evidence for {\it ionization
minima} associated with the radio lobes and hot spots. Such minima are
predicted by the shock models, because the bow shocks driven ahead of
the radio lobes/hotspots will compress the gas and lead to low effective ionization parameters.

\item {\bf Kinematic sub-components.}
Using high resolution spectroscopy it has been possible to
analyse the emission line profiles of several emission line species in 
some detail. Crudely, the emission line components can be divided into
two types: narrow (FWHM $ < 500$ km s$^{-1}$) and broad 
(FWHM $ \sim 800$ -- $1500$ km s$^{-1}$).
In terms of spatial location, the broad components 
are confined to the radio lobes and hotspots (this is particularly striking in the case of PKS2250-41: \pcite{villarmartin99}), whereas
the narrow components are present across the entire emission line
nebula and can extend well {\it beyond} the radio hotspots. Furthermore,
whereas
the broad components have a low ionization state, and contribute much of
the dip in ionization at the position of the radio lobes, the narrow
components have a higher ionization state. 

\item {\bf Sub-component line ratios.} 
By plotting the broad and narrow components separately on diagnostic
diagrams it has been shown that, while the emission line ratios for the 
narrow components are consistent with AGN or shock precursor photoionization
models, the line ratios for the broad components are more consistent with
a pure post-shock cooling spectrum. The emission line
ratios for the broad and narrow components measured in PKS2250-41 are plotted
separately in Figures 4a,d as filled and open triangles
respectively. The separation between the two components is
particularly notable in the
[OIII](5007+4959)/4363 vs. HeII/H$\beta$ diagram, where the
narrow component ratios are consistent with either AGN or shock precursor photoionization,
whereas the broad component ratios are consistent with the shock models, 
but cannot be reconciled with any of the photoionzation
models.
\end{itemize}
Taken together, these data suggest that we have resolved
the shocks kinematically: the narrow components represent the photoionized
precursor emission, while the broad components represent the cooling
post-shock gas. 

Recently a similar approach of combining
kinematic and ionization information has provided clear evidence for shock ionization of the EELR aligned along
the radio axes of high redshift ($z \sim 1$) radio galaxies
(\pcite{best00}).

If the interpretation of the broad component
in terms of the post-shock gas is correct, then we should also expect to detect significant
X-ray continuum and optical coronal line emission from the cooling hot gas (\pcite{wilson99}).
To date the coronal 
component has only been detected in one
EELR in a powerful radio galaxy (PKS2152-69: \pcite{tadhunter88},
Fosbury, private communication). However, with the higher sensitivity
becoming available, it should soon
be possible to detect this component in other objects. The study
of the coronal emission is one of the most direct routes to determining accurate shock parameters.

\section{Conclusions and future work}
\label{sec:discuss}

The emission line nebulae around powerful radio galaxies are complex, but by
using a combination of morphological, kinematical and ionization information
it is now possible to start to distinguish the dominant physical mechanisms.
The strongest result is that there is now compelling evidence that at least
a subset of EELR aligned along the radio axes of powerful radio galaxies
are ionized and accelerated in jet-induced shocks. The main outstanding
questions are as follows.
\begin{itemize}
\item What is the balance between AGN photoionization and shock
ionization (including shock-photoionized precursor) for the EELR that are
not obviously associated with radio features?
\item Why do powerful radio galaxies at high redshifts ($z > 0.6$) appear to show stronger evidence for jet-induced shocks than their lower redshift counterparts?
\item Given that the broad wings to the emission lines are likely to be symptomatic of the entrainment and eventual destruction of the clouds in
the host post-shock wind \cite{villarmartin99}, is jet-induced star formation viable?
\item How important are starburst- or AGN-driven winds in the near-nuclear regions of powerful radio galaxies?
\end{itemize}
Key future observations are likely to include: 
integral field spectroscopy to accurately
measure the velocity shear and ionization 
of the shocked gas relative to the ambient ISM; and deep X-ray imaging and optical spectroscopy to study the hot cooling gas and thereby determine  accurate shock parameters. 

\acknowledgements 
I am grateful to Dave Axon, Neil Clark, Montse Villar-Martin, Andy Robinson, Bob Fosbury,
Raffaella Morganti, Tim Robinson, Anton Koekemoer,
and Carmen Sol\'{o}rzano I\~{n}arrea for their input over
the years. I also thank Carmen for helping to prepare Figure 4, and Anton for
preparing Figure 2b.


\end{document}